\documentstyle[12pt]{article}
\font\titlefont=cmbx10 scaled \magstep3

\def\lprox{\mathrel{\raise .3ex\hbox{$<$\kern-
.75em\lower1ex\hbox{$\sim$}}}}

\def\gprox{\mathrel{\raise .3ex\hbox{$>$\kern-
.75em\lower1ex\hbox{$\sim$}}}}

\begin{document}

\begin{flushright}
\vspace*{-2cm}
gr-qc/9506052 \\ MIT-CTP-2446 \\ TUTP-95-1 \\ Dec. 11, 1995
\vspace*{1.5cm}
\end{flushright}

\begin{center}
{\titlefont AVERAGED ENERGY CONDITIONS \\
\vspace*{0.1in}
AND \\
\vspace*{0.1in}
EVAPORATING BLACK HOLES}\\
\vskip .7in
L.H. Ford\\
\vskip .2in
Center for Theoretical Physics, Laboratory for Nuclear Science\\
Massachusetts Institute of Technology, Cambridge, Massachusetts 02139\\
\vskip 0.1in
and
\vskip 0.1in
Institute of Cosmology, Department of Physics and Astronomy\\
Tufts University, Medford, Massachusetts 02155\footnote{Permanent address;
email: lford@pearl.tufts.edu}\\
\vskip 0.3in
and
\vskip 0.3in
Thomas A. Roman\\
\vskip .2in
Institute of Cosmology, Department of Physics and Astronomy\\
Tufts University, Medford, Massachusetts 02155\\
\vskip 0.1in
and
\vskip 0.1in
Department of Physics and Earth Sciences\\
Central Connecticut State University, New Britain, CT 06050\footnote{Permanent
 address; email: roman@ccsu.ctstateu.edu}\\
\end{center}

\newpage
\begin{abstract}
    In this paper the averaged weak (AWEC) and averaged null (ANEC) energy
conditions, together with uncertainty principle-type restrictions on
negative energy (``quantum inequalities''), are examined in
the context of evaporating
black hole backgrounds in both two and four dimensions. In particular,
integrals over only half-geodesics are studied. We determine the regions of
the spacetime in which the averaged energy conditions are violated. In all
cases where these conditions fail, there appear to be quantum
inequalities which bound the magnitude and extent of the negative energy,
and hence the degree of the violation. The possible relevance of
these results for the validity of singularity theorems in evaporating black
hole spacetimes is discussed.

\end{abstract}
\newpage

\baselineskip=14pt
\section{Introduction}
\label{sec:intro}
   It is by now well-known that quantum field theory permits violations
of all of the local energy conditions used in classical general relativity.
These conditions are employed in a variety of ways, such as in the proofs of
theorems on the occurrence of singularities in gravitational collapse and
cosmology, or of cosmic censorship. Two of the weakest such local conditions
are the ``weak energy condition'' and the ``null energy condition'' which
state that:
\begin{equation}
T_{\mu\nu}\,u^{\mu}\,u^{\nu} \geq 0   \,,              \label{eq:WEC}
\end{equation}
for all timelike vectors $u^{\mu}$, and
\begin{equation}
T_{\mu\nu}\,K^{\mu}\,K^{\nu} \geq 0   \,,              \label{eq:NEC}
\end{equation}
for all null vectors $K^{\mu}$, respectively \cite{HE}. Note that the null
energy condition follows by continuity if the weak energy condition holds.
These two local conditions are satisfied by known forms of classical matter,
but there are a variety of states of quantum fields which violate them
\cite{EGJ}, the most well-known of which is arguably the Casimir vacuum
\cite{C}.

  The extent to which quantum field theory allows violations of the local
energy conditions is not yet completely clear, although progress has been made
in recent years to answer this question. Two principal approaches for
determining the degree of violation have been ``averaged energy conditions''
and ``quantum inequalities.'' The first, originally due to Tipler
\cite{T}, involves a suitable averaging of the local conditions
over timelike or null geodesics. It can be shown that many of the standard
results of classical general relativity obtained via global techniques can be
proved using only the averaged, rather than the local, energy conditions
\cite{T}-\cite{FSW}.
Here we take the ``averaged weak energy condition'' (AWEC) to be
\begin{equation}
\int_{-\infty}^{\infty} \,T_{\mu\nu}\,u^{\mu}\,u^{\nu} \, d\tau \,\geq 0  \,,
                                                          \label{eq:AWEC}
\end{equation}
where the integral is taken over a timelike geodesic with tangent vector
$u^{\mu}$ and parameterized by the proper time $\tau$. Similarly, we take the
``averaged null energy condition'' (ANEC) to be
\begin{equation}
\int_{-\infty}^{\infty} \,T_{\mu\nu}\,K^{\mu}\,K^{\nu} \, d\lambda \,\geq 0 \,,
                                                          \label{eq:ANEC}
\end{equation}
where the integral is taken over an affinely parameterized null geodesic
with tangent vector $K^{\mu}$ and affine parameter $\lambda$
\cite{CONCOM}. There has been
a great deal of effort in the last several years to determine whether
quantum field theory enforces averaged energy conditions. Most of this
activity has been concentrated primarily on ANEC
\cite{K}-\cite{Y95}. This is in part
due to the discovery that violations of ANEC are required to maintain
traversable wormholes \cite{MT,MTY}. It appears that although ANEC
holds for a wide class of quantum states in a variety of spacetimes, it
does not hold in an arbitrary four-dimensional curved spacetime
(see Refs. \cite{WY}, \cite{VISSER}, and \cite{Y94} for more detailed
discussions). However, it is quite plausible that a suitable generalization
of ANEC may hold in more general spacetimes \cite{Y95,FR95}.

The second approach involves uncertainty principle-type inequalities, derived
from quantum field theory, which restrict the magnitude and duration of
negative energy fluxes or densities \cite{FR95,F78,F91}. For
example, one such ``quantum inequality'' (QI) for negative energy fluxes seen
by inertial observers in two-dimensional flat spacetime has the form:
\begin{equation}
 |F| \,(\Delta \tau)^2 \lprox 1 \,,                \label{eq:FI/2D}
\end{equation}
where $|F|$ is the magnitude of the negative energy flux and $\Delta \tau$
is its duration. This inequality implies that $\Delta E$, the amount of
negative energy transmitted in time $\Delta \tau$, is limited by
\begin{equation}
 |\Delta E| \,\Delta \tau \lprox 1 \,.                \label{eq:ETI/2D}
\end{equation}
Therefore, $\Delta E$ is less than the quantum uncertainty in the energy,
${(\Delta \tau)}^{-1}$, on the timescale $\Delta \tau$. Recently, more
precise versions of these inequalities have been derived \cite{F91}. These
involve an integral of the energy flux multiplied by a ``sampling function'',
i.e., a peaked function of time with a time integral of unity and
characteristic width $\tau_0$. A convenient choice \cite{CSAMPLE} of such
a function is
${\tau_0}/[\pi ({\tau}^2+{\tau_0}^2)]$. If the integrated flux, $\hat F$,
is defined by
\begin{equation}
\hat F \equiv { {\tau_0} \over {\pi} } \,
\int_{-\infty}^{\infty}\, { {F(\tau)} \over { {\tau}^2 + {\tau_0}^2 } } \,
d\tau \,,                                    \label{eq:def-FHAT}
\end{equation}
then these inequalities may be written as
\begin{equation}
\hat F \gprox - {1 \over {16 \pi {\tau_0}^2} } \,,    \label{eq:FQI/2D}
\end{equation}
and
\begin{equation}
\hat F \gprox - {3 \over {32 {\pi}^2 \,{\tau_0}^4} } \,,  \label{eq:FQI/4D}
\end{equation}
for all $\tau_0$, in two- and four-dimensions, respectively. These inequalities
have the form required to prevent macroscopic violations of the second law of
thermodynamics \cite{F78,F91}. It was also discovered that similar
inequalities hold for a quantized massless, minimally-coupled scalar field
propagating on two- and four-dimensional extreme Reissner-Nordstr{\"o}m
black hole backgrounds. These inequalities were shown to foil attempts to
produce an unambiguous violation of cosmic censorship by injecting
a negative energy flux into an extreme charged black hole \cite{FR90,FR92}.
The latter results provide perhaps one of the strongest reasons for the
belief that the production of large-scale effects via manipulation of
negative energy may be forbidden by quantum field theory. Classically,
{\it any} amount of negative energy, no matter how small, injected into
an extreme black hole is enough to produce a naked singularity. Therefore,
if one had any chance of producing gross effects with negative energy, the
scenario discussed above would seem to have offered the best
possibility of success. It should also be emphasized that the energy-time
uncertainty
principle was {\it not} used as input in the derivation of any of the
QI restrictions. They arise directly from quantum field theory.

In a recent paper, we have shown that there exist deep connections between
averaged energy conditions and quantum inequalities, at least in flat
spacetime \cite{FR95}. For a quantized massless, minimally-coupled scalar
field in two- and four-dimensional Minkowski spacetime, we have derived
analogous QI's to Eqs.~(\ref{eq:FQI/2D}) and ~(\ref{eq:FQI/4D}) for energy
density. In particular, for timelike geodesics in two-dimensional flat
spacetime,
\begin{equation}
{\tau_0 \over \pi} \int_{-\infty}^{\infty}
{{\langle T_{\mu\nu}u^\mu u^\nu\rangle d{\tau}}\over {{\tau}^2+{{\tau}_0}^2}}
\geq -{1\over {8\pi {{\tau_0}^2}}},                    \label{eq:TLQI2DUNCP}
\end{equation}
for all $\tau_0$. Here $\langle T_{\mu\nu}u^\mu u^\nu\rangle$ is the
renormalized expectation value taken in an arbitrary quantum state
$\mid\psi\rangle$. In the limit as $\tau_0 \rightarrow \infty$, we
``sample'' the entire geodesic, and our inequality
Eq.~(\ref{eq:TLQI2DUNCP}) reduces to AWEC. Note that it is possible that the
integral in Eq.~(\ref{eq:TLQI2DUNCP}) converges, even though the associated
AWEC integral in Eq.~(\ref{eq:AWEC}) diverges. Analogous results hold in this
case for null geodesics.

For a quantized massless scalar field in a two-dimensional Minkowski
spacetime with compactified spatial dimension,
it was also discovered that the {\it difference} between the expectation values
of $T_{\mu\nu} u^{\mu} u^{\nu}$, or $T_{\mu\nu} K^{\mu} K^{\nu}$, in an
arbitrary quantum state and in the Casimir vacuum state obey AWEC and
ANEC-type inequalities \cite{FR95}. This is surprising since it is
known \cite{K} that AWEC and ANEC are violated for
$\langle T_{\mu\nu} u^{\mu} u^{\nu} \rangle$ and
$\langle T_{\mu\nu} K^{\mu} K^{\nu} \rangle$, respectively, in the
(renormalized) Casimir vacuum state by itself. Such ``difference inequalities''
might provide new measures of the degree of energy condition violation
in cases where the usual averaged energy conditions
fail. This approach has recently been generalized to arbitrary two-dimensional
curved spacetimes by Yurtsever \cite{Y94}.

For a quantized massless, minimally-coupled scalar field in
four-dimensional Minkowski spacetime, we also derived the following
inequality for timelike geodesics
\begin{equation}
\hat \rho = {{\tau_0} \over \pi}\, \int_{-\infty}^{\infty}\,
{{\langle T_{\mu\nu} u^{\mu} u^{\nu}\rangle\, d\tau}
\over {{\tau}^2+{\tau_0}^2}} \geq
-{3\over {32 {\pi}^2 {\tau_0}^4}}\,,  \label{eq:4DENQI}
\end{equation}
for all $\tau_0$. From this inequality, it was shown that one can
derive AWEC and ANEC \cite{FR95}.
(The application of QI bounds to constrain the dimensions of traversable
wormholes is discussed in a separate manuscript\cite{FR-Worm}.)

In almost all studies of ANEC for quantum fields, the bounds on the
ANEC integral have been taken from $-\infty$ to $+\infty$
\cite{K,WY,Y}. However, a question of some interest is
whether ANEC is satisfied in the spacetime of
an object collapsing to form a black hole. The answer to this question
might determine, for example, whether Penrose's
singularity theorem \cite{P65} will still hold in the presence of
local violations of the energy conditions, such as the Hawking evaporation
process \cite{H75,B-TC}. In this case, one is usually
concerned with the focusing of null geodesics which generate the
boundary of the future of a trapped surface. To prove Penrose's theorem,
one would want ANEC to hold over these half-complete geodesics
\cite{TR-86,TR-88}. This is a stronger condition to impose than to
demand that ANEC hold over full (i.e., past and future-complete) geodesics.
However, even if this version of ANEC fails in some regions of a given
spacetime, it may hold in enough other regions so that the conclusions of
Penrose's theorem will still be valid. In addition, it should be noted
that ANEC is a sufficient condition to insure continued focusing of null
geodesics. One can also guarantee focusing with other conditions, for
example by allowing the right-hand-side of the ANEC integral to be only
periodically non-negative \cite{B87}, or even negative but bounded
(see Lemma 3 of Ref. \cite{Galloway},  Eq. (5) of Ref. \cite{TR-88},
and Ref. \cite{Y95}).

In the current paper, we examine quantized massless scalar, and
electromagnetic fields in the Unruh vacuum state in two- and
four-dimensional Schwarzschild black hole backgrounds. We evaluate
AWEC and ANEC integrals along half-complete timelike
and null geodesics, respectively. Our goal is to determine the regions of
these spacetimes in which the averaged energy conditions fail, and whether
there exist any bounds on the extent of the violation.
The paper is organized as follows. The analysis of 2D evaporating black
holes is presented in Sec.~(\ref{sec:2DEVAP}); the results for 4D black
holes are given in Sec.~(\ref{sec:4DBH}). The latter section also includes
a discussion of ANEC along orbiting null geodesics. Concluding remarks are
given in Sec.~(\ref{sec:CONC}). Our units are taken to be: $\hbar=G=c=1$.

\section{2D Evaporating Black Holes}
\label{sec:2DEVAP}
       In this section, we study AWEC and ANEC for a quantized,
massless scalar field in the Unruh and Boulware vacuum states, along
geodesic segments in a 2D Schwarzschild spacetime. The discussion here
is a natural extension of the analysis in Ref. \cite{FR93}.

\subsection{Timelike Observers}

          Let $u^\mu$ be the two-velocity of an inertial observer.
The energy density in this observer's frame is given by
\begin{equation}
U = T_{\mu\nu} {u^\mu} {u^\nu}\, .    \label{eq:enden}
\end{equation}
In the following discussion, we understand
$T_{\mu\nu}$ to denote a quantum expectation
value in a specified vacuum state. The metric is
\begin{equation}
ds^2 = -C dt^2 + C^{-1} dr^2,            \label{eq:2dmetric}
\end{equation}
where $C = 1 - 2M/r$. An outgoing geodesic observer's two-velocity is
\begin{equation}
u^\mu = (u^t, u^r) = \Bigl({{dt} \over {d\tau}}, {dr\over {d\tau}} \Bigr) =
        \Bigl( {k \over C},  \sqrt{k^2 -C} \Bigr). \label{eq:umu}
\end{equation}
The constant $k$ is the energy per unit rest mass.
     In our two-dimensional discussion, we will consider observers moving
in both the Unruh and Boulware vacua. The stress tensor components in the
Unruh vacuum are \cite{Unruh77}:
\begin{equation}
T_{tt}={1\over{24\pi}} \biggl({{7M^2}\over{r^4}}-{{4M}\over{r^3}}
       +{1\over{32M^2}}\biggr)\,,    \label{eq:UTtt}
\end{equation}
\begin{equation}
T_{tr}= -{1\over{24\pi}} \biggl(1-{{2M}\over r}\biggr)^{-1}{1\over{32M^2}}\,,
\label{eq:UTtr}
\end{equation}
and
\begin{equation}
T_{rr}=-{1\over{24\pi}} \biggl(1-{{2M}\over r}\biggr)^{-2}
        \biggl({{M^2}\over{r^4}}-{1\over{32M^2}}\biggr)\,. \label{eq:UTrr}
\end{equation}
The corresponding components in the Boulware vacuum are given by:
\begin{equation}
T_{tt}={1\over{24\pi}} \biggl({{7M^2}\over{r^4}}-{{4M}\over{r^3}}
       \biggr)\,, \label{eq:BTtt}
\end{equation}
\begin{equation}
T_{tr}=0\,,  \label{eq:BTtr}
\end{equation}
and
\begin{equation}
T_{rr}=-{1\over{24\pi}}{{M^2}\over{r^4}} \biggl(1-{{2M}\over r}\biggr)^{-2}\,.
\label{eq:BTrr}
\end{equation}

For an outgoing timelike observer in the Unruh vacuum,
\begin{equation}
T_{\mu\nu} {u^\mu} {u^\nu}= {1\over {24\pi}}\,C^{-2}\, \Biggl\{
k^2\,\Biggl[ {{6M^2} \over{r^4}} - {4M\over {r^3}} +
{1\over {16 M^2}} \Biggr]\,
+\,{{CM^2} \over {r^4}} \,- \, {C\over {32 M^2}}\, - \,
{{k \sqrt{k^2-C}} \over {16 M^2}} \Biggr\}.                  \label{eq:etl/out}
\end{equation}
Let us consider, for simplicity, the case
of an observer who starts out initially very close to the horizon and is shot
outward at large velocity, i.e., we are interested in the limits
\begin{equation}
\epsilon \ll M\, , \qquad k \gg 1 \,,      \label{eq:k-lim}
\end{equation}
where $r = 2M + \epsilon$ is the observer's initial position. In this limit
$k \gg C$, so from the geodesic equation, Eq.~(\ref{eq:umu}),
we have that
\begin{equation}
\tau \sim \frac{r-2M-\epsilon}{k} + O(k^{-3})\,.   \label{eq:tauexp}
\end{equation}
We now wish to
multiply $T_{\mu\nu} {u^\mu} {u^\nu}$ by a sampling function, i.e., a
peaked function of time whose time integral is unity. In Refs. \cite{FR95}
and \cite{F91}, this function was chosen to be
\begin{equation}
{\tau_0 \over \pi} \int_{-\infty}^{\infty}
\,{{d\tau} \over {{\tau}^2 +{\tau_0}^2}}\, =\,1\,,      \label{eq:origsamp}
\end{equation}
where the integral was taken over complete geodesics.
However, unlike in Ref. \cite{FR95}, here we are integrating over
half-infinite geodesics. Since the function given by
Eq.~(\ref{eq:origsamp}) is symmetric about $\tau=0$, in our case we may choose
\begin{equation}
{ {2\tau_0} \over \pi} \int_{0}^{\infty}
\,{{d\tau} \over {{\tau}^2 +{\tau_0}^2}}\, =\,1\,,         \label{eq:samp}
\end{equation}
where the proper time $\tau$ is initialized when the observer starts at
$r=2M+\epsilon$. We multiply $T_{\mu\nu} {u^\mu} {u^\nu}$ by
this sampling function, use Eq.~(\ref{eq:tauexp}), and expand the numerator
in inverse powers of $k$. The integral is performed using MACSYMA and the
result is expanded in powers of $\epsilon$, yielding
\begin{eqnarray}
{ {2\tau_0} \over \pi} \int_{0}^{\infty}
\,{{T_{\mu\nu} {u^\mu} {u^\nu} d\tau} \over
{{\tau}^2 +{\tau_0}^2}}\, & \sim &
-\frac{k}{24 \pi^2\,\tau_0} \, \biggl[ {1 \over {\epsilon} } +
 {1 \over M} \, \ln \biggl({\epsilon \over 2M} \biggr)
+ O({\epsilon}^0) \biggr]             \nonumber \\
& + & \frac{M}{3 \pi^2 \,{\tau_0}^3 \,k} \, \ln \biggl(
\frac{k \tau_0}{2M} \biggr) \, + O(k^{-1}) \,.   \label{eq:intiT}
\end{eqnarray}
Therefore, to leading order in $k$ and $\epsilon$, we have
\begin{equation}
{ {2\tau_0} \over \pi} \int_{0}^{\infty}
\,{{T_{\mu\nu} {u^\mu} {u^\nu} d\tau} \over
{{\tau}^2 +{\tau_0}^2}}\,
 \sim  -{1 \over {24{\pi}^2 \, \tau_0 \,\delta \tau}}\,,  \label{eq:intT}
\end{equation}
where
\begin{equation}
{\delta \tau} = {{\epsilon} \over k}\,.           \label{eq:deltau}
\end{equation}
Note that in these limits $\delta \tau \ll M$ and
$\delta \tau \ll \tau_0$ \cite{2TS}.

What is the physical significance of $\delta \tau$?
The negative energy density drops off very rapidly with increasing $r$.
In the limit $k \gg 1$ and $\epsilon \ll M$, if we
consider the proper time spent by the observer in the region of
appreciable negative energy to be
\begin{equation}
\int_{0}^{\delta \tau} \, d\tau \, \sim \, {1 \over k} \,
\int_{2M + \epsilon}^{2M + x \epsilon} \, dr \, =
                             {1 \over k} \, \epsilon(x -1) \,,
                                           \label{eq:in/deltatau}
\end{equation}
then typically $x-1 \sim O(1)$, so
\begin{equation}
{\delta \tau} \sim {{\epsilon} \over k} \,.       \label{eq:lim/deltau}
\end{equation}
To see this a little more explicitly, note that near the horizon,
Eq.~(\ref{eq:etl/out}) becomes
\begin{equation}
T_{\mu\nu} {u^\mu} {u^\nu} \sim -{{k^2} \over {48\pi\Bigl( r-2M \Bigr)}^2}
\,\, . \label{eq:Tasyp}
\end{equation}
If the observer starts at
$r=r_0= 2M + \epsilon$, then at $r=r_1= 2M + 2\epsilon$,
\begin{equation}
\Bigl( T_{\mu\nu} {u^\mu} {u^\nu} \Bigr)|_{r=r_1} \approx
{1 \over 4} \,\Bigl( T_{\mu\nu} {u^\mu} {u^\nu} \Bigr)|_{r=r_0}\,,
\end{equation}
so the energy density drops off to $1/4$ of its initial value in a
distance $\Delta r = \epsilon$, corresponding to a proper time
$\delta \tau = \epsilon/k$. Therefore, we should consider the time
interval $\delta \tau$ not to be the entire proper time over which the energy
density is negative, but rather a time scale which characterizes a change in
the energy density.

For an outgoing timelike observer in the Boulware vacuum:
\begin{equation}
T_{\mu\nu} {u^\mu} {u^\nu}= {1\over {24\pi}}\,C^{-2}\, \Biggl\{
k^2\,\Biggl[ {{6M^2} \over{r^4}} - {4M\over {r^3}} \Biggr]\,
+\,{{CM^2} \over {r^4}} \Biggr\}.                  \label{eq:betl/out}
\end{equation}
Note that this quantity is negative everywhere for $r>2M$. If we perform
the same set of operations as discussed above, we find that
Eq.~(\ref{eq:intT}) holds for the Boulware vacuum as well. In
the $\tau_0\rightarrow \infty$ limit of Eq.~(\ref{eq:intT}),
we ``sample'' the entire (i.e., half-infinite) geodesic and find
\begin{equation}
\int_{0}^{\infty}\, T_{\mu\nu} {u^\mu} {u^\nu} d\tau \approx
-{1 \over {48\pi \delta \tau}}\,,     \label{eq:bawec/2d}
\end{equation}
and hence
\begin{equation}
\int_{0}^{\infty}\, T_{\mu\nu} {u^\mu} {u^\nu} d\tau \gprox
-{1 \over {\delta \tau}}\,.                            \label{eq:awec/qi}
\end{equation}
The dominant contribution to the integral in Eq.~(\ref{eq:bawec/2d})
is from the lower limit, since $T_{\mu\nu} {u^\mu} {u^\nu}$ in the Boulware
vacuum state also has the form given in Eq.~(\ref{eq:Tasyp}) near the horizon,
but drops off rapidly with increasing $r$. By contrast, the
$\tau_0\rightarrow \infty$ limit of Eq.~(\ref{eq:intT}) for the Unruh
vacuum state results in an integral which diverges positively at large $r$.
This is due to the fact that in 2D the Hawking radiation does not drop
off with distance, but remains constant at large $r$. Therefore, AWEC is
satisfied in the Unruh vacuum state in 2D. However, as discussed in
Sec.~(\ref{sec:4DBH}), the situation is different in 4D.
We could perform the integral on the left-hand-side of
Eq.~(\ref{eq:bawec/2d}) for the Unruh vacuum state, but instead truncate
the integration at some finite value of $r=r_{max}$, with
$2M + \epsilon < r_{max} < \infty$. For fixed $r_{max}$, by making $\epsilon$
small enough we can always arrange it so that the negative energy
contribution from the lower limit, given by Eq.~(\ref{eq:Tasyp}), dominates
the integral. In that case, using Eq.~(\ref{eq:Tasyp}), we have that
\begin{equation}
\int_{0}^{\tau(r_{max})}\, T_{\mu\nu} {u^\mu} {u^\nu} d\tau \approx
-{1 \over {48\pi \delta \tau}}\, \gprox
-{1 \over {\delta \tau}}\,.                            \label{eq:uawec/qi}
\end{equation}

The inequalities, Eqs.~(\ref{eq:awec/qi}) and ~(\ref{eq:uawec/qi}),
represent bounds on the degree
of AWEC violation seen by timelike geodesic observers who start out
very close to the horizon and are shot outward at high speed.
In this limit, the longer the (proper) timescale over which the observer
sees a significant change in the energy density, the smaller is the
magnitude of the integrated negative energy density seen by that observer.
Since the negative energy density drops off rapidly with increasing $r$,
to remain in the negative energy density the observer must stay close
to the horizon. The closer the observer is to the horizon, the larger
is the magnitude of the negative energy density.
However, in order to remain close to the horizon for a long time as seen
by a distant observer, the observer's trajectory must be nearly lightlike.
Therefore, although the observer spends a long time in the negative energy
region as seen by the distant observer, the {\it proper} time spent
in the region of appreciable negative energy (as measured by
$\delta \tau$) decreases with the observer's proximity to the horizon.
These inequalities are similar in form to Eq.~(\ref{eq:ETI/2D}).

\subsection{Null Geodesics in 2D: Off the Horizon}

  We now wish to examine ANEC for null geodesics in 2D Schwarzschild
spacetime. From Eq.~(\ref{eq:2dmetric}), and from the equation for null
geodesics, we have
\begin{equation}
{dt \over {d\lambda}}  =   \pm C^{-1} \, E \,, \quad         \label{eq:ngeod}
{dr \over d\lambda}  =  \pm E \,,
\end{equation}
where $E$ is an arbitrary positive constant, whose value fixes the scale of
the affine parameter, $\lambda$. We want to examine ANEC along both ingoing and
outgoing future-directed (i.e., $d\lambda>0$) null geodesics, inside and
outside the horizon. There are four cases (see Fig. 1):
\begin{eqnarray}
1)\,\,{dt \over {d\lambda}} & > & 0\,,\quad {dr \over {d\lambda}} =\,\,\,
E > 0\,,\quad
\,\,\,{\rm for } \, r>2M \,{\rm (outgoing\,\, rays) }\,,
\label{eq:out/rg2M}   \\
2)\,\,{dt \over {d\lambda}} & > & 0 \,,\quad {dr \over {d\lambda}} =
-E < 0\,,\quad
{\rm for } \, r>2M\, {\rm (ingoing\,\, rays) }\,, \label{eq:in/rg2M} \\
3)\,\,{dt \over {d\lambda}} & > & 0\,,\quad {dr \over {d\lambda}} =
-E < 0\,,\quad
{\rm for } \, r<2M \, {\rm (outgoing\,\, rays) }\,, \label{eq:out/rl2M} \\
4)\,\,{dt \over {d\lambda}} & < & 0\,,\quad {dr \over {d\lambda}} =
-E < 0\,,\quad
{\rm for }\, r<2M \, {\rm (ingoing\,\, rays) }\,.      \label{eq:in/rl2M}
\end{eqnarray}
Note that
$C=(1-2M/r)$ changes sign inside the horizon, while $d\lambda$ remains
positive for future-directed null rays.

If we use Eqs.~(\ref{eq:out/rg2M}) - ~(\ref{eq:in/rl2M}), and
Eqs.~(\ref{eq:UTtt}) - (\ref{eq:UTrr}), we obtain for the Unruh vacuum
state
\begin{equation}
T_{\mu\nu} {K^\mu} {K^\nu} =
{E^2\over {24\pi}} \,
{\Biggl(1-{2M \over r}\Biggr)}^{-2}
\,\Biggl[ {{6M^2}\over{r^4}} - {4M\over {r^3}} \Biggr] \,
{\rm (outgoing)}\,,  \label{eq:enull/o}
\end{equation}
and
\begin{equation}
T_{\mu\nu} {K^\mu} {K^\nu} =
{E^2\over {24\pi}} \,
{\Biggl(1-{2M \over r}\Biggr)}^{-2}
\,\Biggl[ {{6M^2} \over{r^4}} - {4M\over {r^3}} + {1\over {8 M^2}} \Biggr]
{\rm (ingoing)} \,.                 \label{eq:enull/i}
\end{equation}
To obtain Eq.~(\ref{eq:enull/i}), we have used the fact that the sign of
$T_{tr}$ changes inside the horizon, as can be seen from Eq.~(\ref{eq:UTtr}),
together with Eqs.~(\ref{eq:in/rg2M}) and ~(\ref{eq:in/rl2M}).
Note that for Eq.~(\ref{eq:enull/i}), the term in square brackets vanishes
as $r\rightarrow 2M$. A Taylor expansion shows that, to leading order, it
vanishes as ${(r-2M)}^2$, so it will cancel the ${(r-2M)}^2$ divergence
in the denominator. Therefore Eq.~(\ref{eq:enull/i}) is finite on the
horizon, as it should be. However, there appears to be a discontinuity in
Eq.~(\ref{eq:enull/o}) for an outgoing null ray on the horizon, $r=2M$.
The situation for null rays on the horizon is sufficiently subtle as to
warrant a separate discussion. This is provided in the next subsection.

{}From Eq.~(\ref{eq:enull/o}), $T_{\mu\nu} {K^\mu} {K^\nu}$ is negative for
outgoing null geodesics when\linebreak $r>1.5 M$.
Therefore the local null energy condition, Eq.~(\ref{eq:NEC}) is violated
along outgoing null vectors slightly inside the horizon, as well as along
all outgoing rays outside (and on) the horizon. For $r>2M$, consider
an outgoing null geodesic starting at $r=2M+\epsilon$, with
$\epsilon \ll M$. Examine ANEC along this ray to obtain
\begin{equation}
I_1 \, \equiv \,\int_{0}^{\infty}
\,{T_{\mu\nu} {K^\mu} {K^\nu} d\lambda} =
\,{E \over {24\pi}} \, \int_{2M+\epsilon}^{\infty}
\, {1 \over {\Bigl( r-2M \Bigr)}^2}
\,\Biggl[ {{6M^2}\over{r^2}} - {4M\over {r}} \Biggr]\,dr\,,
                                       \label{eq:ANEC/2D out/rg2M}
\end{equation}
where we have used Eq.~(\ref{eq:out/rg2M}). Here $\lambda=0$ is the
value of the affine parameter at $r=2M+\epsilon$.
Equations~(\ref{eq:enull/o}) and ~(\ref{eq:ANEC/2D out/rg2M})
apply for the Boulware vacuum as well as for the Unruh vacuum.
This is due to the fact that in the latter case, the
outgoing null geodesics are parallel to the Hawking radiation. Hence the
terms which would cause the integral in Eq.~(\ref{eq:uawec/qi}) to diverge
as $r_{max} \rightarrow \infty$ are absent in the null geodesic case.

For $r<2M$, along an outgoing
null geodesic starting just inside the horizon at $r=2M-\epsilon$:
\begin{equation}
I_2 \, \equiv \,\int_{0}^{\lambda_f}
\,{T_{\mu\nu} {K^\mu} {K^\nu} d\lambda} =
\,{E \over {24\pi}} \, \int_{r_{min}}^{2M-\epsilon}
\, {1 \over {\Bigl( r-2M \Bigr)}^2}
\,\Biggl[ {{6M^2} \over{r^2}} - {4M\over {r}} \Biggr]\,dr\,,
                                           \label{eq:ANEC/2D out/rl2M}
\end{equation}
where we have used Eq.~(\ref{eq:out/rl2M}), and $\lambda_f$ is the
value of the affine parameter at $r=r_{min}$, the minimum value of $r$
attained by the null geodesics.
Notice that the integrand in
Eq.~(\ref{eq:ANEC/2D out/rl2M}) will be dominated by the positive
$6M^2/{r^2}$ term which diverges for small $r$. From Eq.~(\ref{eq:enull/i}),
we see that $T_{\mu\nu} {K^\mu} {K^\nu}$ is positive everywhere for ingoing
null geodesics. Thus we immediately see that ANEC is satisfied for these
two sets of null geodesics.

In Eq.~(\ref{eq:ANEC/2D out/rg2M}) we perform the integration and then
expand the result in the limit of small $\epsilon$ to find
\begin{equation}
\lim_{\epsilon \rightarrow 0}\, I_1 \, \sim
-{E \over {48\pi \,\epsilon}} + O\bigl(\log(\epsilon/M)\bigr) <\,0 \,.
       \label{eq:lim out/rg2M}
\end{equation}
Note that $\delta \lambda = E/\epsilon$ is the characteristic affine parameter
distance over which the negative energy is decreasing along the outgoing
geodesic. We can rewrite Eq.~(\ref{eq:lim out/rg2M}) as
\begin{equation}
\int_{0}^{\infty}\, T_{\mu\nu} {K^\mu} {K^\nu} d\lambda \approx
-{1 \over {48\pi \delta \lambda}}\, \gprox
-{1 \over {\delta \lambda}}\,.       \label{eq:anec/nint}
\end{equation}
This is the null version of Eq.~(\ref{eq:uawec/qi}). Note that it is invariant
under rescaling of the affine parameter.

Similarly, for Eq.~(\ref{eq:ANEC/2D out/rl2M}),
we may perform the integration and expand the result in the small $r_{min}$
limit. For fixed $\epsilon$, the result is
\begin{equation}
\lim_{r_{min} \rightarrow 0}\, I_2 \, \sim {{E} \over {16\pi \, r_{min}}} \,
+ O\bigl(\ln(r_{min}/M)\bigr) \,
>0  \,. \label{eq:lim out/rl2M}
\end{equation}
However, for fixed $r_{min}$ and $\epsilon \rightarrow 0$, $I_2$ has the same
behavior as $I_1$ :
\begin{equation}
 I_2 \, \sim
-{E \over {48\pi \,\epsilon}} + O\bigl(\ln(\epsilon/M)\bigr) <\,0 \,.
       \label{eq:lim2 out/rl2M}
\end{equation}

Thus we see that ANEC is violated for outgoing null rays just outside the
horizon, but it is satisfied for outgoing null rays inside the horizon
in the limit $r_{min} \rightarrow 0$.
(The divergence in Eq.~(\ref{eq:lim out/rg2M}) as $\epsilon \rightarrow 0$
may be circumvented by an appropriate choice of scaling of the affine
parameter, i.e., by an appropriate choice of $E$. This and related issues will
be discussed in the next sub-section.) Outgoing rays which
originate in the region $1.5M<r<2M$ initially encounter negative energy,
and later encounter positive energy near the singularity, $r=0$. This is
analogous to the case of outgoing timelike geodesics in the Unruh vacuum,
which encounter negative energy near the horizon and positive energy at
large distances due to the Hawking radiation.
 We also see that ANEC holds for ingoing null geodesics, whether they
originate outside or inside the horizon.

One might at first sight conclude that the reasoning here is a bit circular.
It could be argued that it is no surprise that evaluation of the
ANEC integral for null geodesics inside the horizon yields a divergent
positive result. After all, we placed our quantized field on a background
which already had a singularity in it. Thus one could argue that
divergent values of our
ANEC integrals for these geodesics are not unexpected, because the behavior
of the fields inside the horizon is dominated by the singularity.
However, it is
not obvious, a priori, what the {\it sign} of these integrals should be.
Since these are test fields on a given background, as opposed to fields
which generate the background spacetime (which we cannot have in any case
in 2D), it is possible that the integrals might have turned out to
diverge with either sign.
Similar ANEC integrals for quantized massless fermion fields in 2D have the
same signs as for the massless scalar field. This follows from the fact that
the renormalized fermion stress tensor is identical to that in the scalar case
\cite{DU77}, even though the formally divergent tensors have opposite signs in
the two cases.

\subsection{Null Geodesics in 2D: On the Horizon}

  Recall that our results for outgoing null geodesics outside the horizon are
formulated in terms of the null tangent vector
\begin{equation}
K^{\mu} = E \, ( C^{-1},\, 1) \,.       \label{eq:snullv}
\end{equation}
where the components are given in terms of Schwarzschild $(t,r)$ coordinates.
Since these coordinates are badly-behaved on the horizon, let us switch to
Kruskal null coordinates, given by
\begin{equation}
U=-e^{-\kappa u}\,, \quad V=e^{\kappa v}\,,   \label{eq:Kcoords}
\end{equation}
where $u=t-r^{\ast}$, $v=t+r^{\ast}$, with $r^{\ast}$ the usual tortoise
coordinate and $\kappa= 1/(4M)$. In these
coordinates, our null vector has the form
\begin{equation}
K^{\bar \mu} = \Biggl( \frac{dU}{d\lambda}, \frac{dV}{d\lambda} \Biggr)
 = \Biggl(0\,, \frac{2E}{C} \, \kappa \, e^{\kappa v} \Biggr)\,.
                                                   \label{eq:Knullv}
\end{equation}
If $E$ has the same value for all outgoing null geodesics, then $K^{\bar \mu}$
would not be defined in the limit where the geodesics approach the horizon,
since $C \rightarrow 0$. However, $E$ need only be constant along each
null geodesic, but not the same constant for different null geodesics.

Consider a sequence of outgoing null geodesics which start at different values
of $r$, e.g., the histories of a sequence of photons emitted by an infalling
observer. Let $r_0$ be the value of $r$ at which the null geodesic begins
(which will be different for each null geodesic). We might choose the
various values of $E$ for
different null geodesics by the following prescription. Let the observer emit
photons of fixed frequency $\omega_e$ in his rest frame. The time-component
of the observer's two-velocity (in Schwarzschild coordinates) is
\begin{equation}
u^t=\,{ {dt} \over {d\tau} } \, = \,
{ \Biggl(1- { {2M} \over {r_0} }\Biggr) }^{-1} \, {\tilde E}\,,
                                               \label{eq:2v-in}
\end{equation}
where ${\tilde E}$ is the observer's energy per unit rest mass, measured
at infinity. This is just
the rate of a clock at infinity as compared to the rate of the clock carried
by the infalling observer. The frequency of a photon at infinity is
$\omega_{\infty}$, which is related to $\omega_e$ by
\begin{equation}
{ {\omega_{\infty}} \over {\omega_e} }\, = \, { {d\tau} \over {dt} }\,
\,= \,  \Biggl(1- { {2M} \over {r_0} }\Biggr)  \, {\tilde E}^{-1}\,.
                                                      \label{eq:freqrat}
\end{equation}
Note that the expression on the right-hand-side includes both the Doppler
effect and the gravitational redshift. The factor of ${\tilde E}^{-1}$
reflects the fact that the faster the observer is shot toward
the black hole (i.e., the larger ${\tilde E}$ is), the greater will
be the Doppler shift. If we choose a scaling for the affine parameter such that
$E= \omega_{\infty}$, for fixed $\omega_e$, then
\begin{equation}
E\,=\, \Biggl(1- { {2M} \over {r_0} }\Biggr)  \, \omega_e \, {\tilde E}^{-1}
\, \propto C(r_0)\,.                                    \label{eq:EpropC}
\end{equation}
Now $E/{C(r_0)}$ is finite in the limit $r_0 \rightarrow 2M$, so that the
components of $K^{\mu}$ are finite on the horizon, in either
Schwarzschild or Kruskal coordinates.

However, $K^{\bar \mu}$ is still not the same as $k^{\bar \mu}$, the
{\it affinely}
parameterized null tangent vector on the horizon. In Kruskal null coordinates
the vector $k^{\bar \mu}$ has components
\begin{equation}
k^{\bar \mu}\,=\, (0,\, \kappa)\,.            \label{eq:def-k}
\end{equation}
(See the discussion on p. 331 of Ref. \cite{WALDGR}.)
On the horizon,
\begin{equation}
K^{\bar \mu}\,=\, {\Biggl( { {2E} \over C } \Biggr)}_{r=2M} \! e^{\kappa v} \,
(0\,, \kappa)\,,                                \label{eq:K-k/Krus}
\end{equation}
so these vectors still differ by a factor of ${\rm exp}(\kappa v)$. Thus,
a null vector such as $K^{\mu}$, which is an affinely parameterized null
tangent vector off the horizon, does not necessarily remain affinely
parameterized {\it on} the horizon.

With these lessons in hand, let us now consider the integral
$\int_{}^{}\, T_{\bar \mu \bar \nu} \, k^{\bar \mu}\,k^{\bar \nu}\, dV$
along a portion of the future event horizon. Here $T_{\bar \mu \bar \nu}$
is the vacuum expectation value of the stress-tensor in the Unruh vacuum state
expressed in Kruskal null coordinates, and $k^{\bar \mu}$ is an affinely
parameterized null tangent vector on the horizon. The Kruskal
advanced time coordinate $V$, is
an affine parameter on the horizon (see for example, p. 331 of
Ref. \cite{WALDGR} or p. 122 of Ref. \cite{WALDQFT}). From
Eq.~(\ref{eq:def-k}), and a straightforward coordinate transformation of
Eqs.~(\ref{eq:UTtt})-~(\ref{eq:UTrr}), we obtain
\begin{equation}
\int_{V_0}^{\infty}\, T_{\bar \mu \bar \nu} \, k^{\bar \mu}\,k^{\bar \nu}\, dV
\,= \, - { 1 \over {768\,\pi\,M^2\,V_0} }\,.             \label{eq:Kint}
\end{equation}
This integral is taken over a portion of the future horizon; $V_0$ is the
value of the $V=\,const$ line which intersects the future horizon at the
(arbitrary) event where we start our null geodesic.

The right-hand-side of Eq.~(\ref{eq:Kint}) goes to $-\infty$ as
$V_0 \rightarrow 0$. However, this would correspond to integrating along the
full future horizon of an eternal black hole spacetime. The Unruh vacuum
state is singular on the past horizon (i.e., $V=0$) of such a spacetime,
so it is not surprising that we get a divergence in this case. Physically,
this would correspond to a black hole that Hawking-radiates for an infinite
time. For the Unruh vacuum state to be realizable, we must consider only
physically realistic collapse spacetimes in which the black hole forms at
a finite time in the past. Therefore, the smallest allowed value of
$V_0=V_{min}$, which is the value of the $V=\,const$ line at which the
the worldline of the surface of the collapsing star intersects the
future event horizon, as shown in Fig. (2). For such spacetimes, the
ANEC integral along a portion of the future horizon is negative but finite.
This result seems to be in agreement with the results of Wald and Yurtsever
\cite{WY}. They find that ANEC is satisfied along the complete null geodesic
comprised of the future horizon plus the null line which would have been the
past horizon in an eternal black hole spacetime. In a collapse
spacetime, this complete null geodesic originates at past null infinity,
propagates through the collapsing star,
``reflects off'' $r=0$, and then emerges into the vacuum outside the star
at $V_0=V_{min}$ (see Fig. (2) ). Evidently there is a positive contribution
to the ANEC integral from the part of the geodesic prior to its exit from the
collapsing star \cite{Y-PRIVATE}. This contribution must be greater than or
equal to
$(768\,\pi\,M^2\,V_{min})^{-1}$. (Note that the source of the positive
contribution we refer to here is the quantized field, not the classical
collapsing matter. The latter would presumably make the ANEC integral
even more positive.)

   The dimensions of $k^{\bar \mu}$ are ${(length)}^{-1}$, whereas
$T_{\bar \mu \bar \nu}$, $V$, and $V_0$ are dimensionless. Let us now
rescale the coordinates to get a more familiar choice of dimensions.
Let $x^{\hat \mu}=\,{\kappa}^{-1}\,x^{\mu}$, i.e.,
$\hat U = {\kappa}^{-1}\,U\,, \hat V = {\kappa}^{-1}\,V$, so
$\hat {V_0} = {\kappa}^{-1}\,V_0$. Then $T_{\hat \mu \hat \nu} =
{\kappa}^2 \, T_{\bar \mu \bar \nu}$, and $k^{\hat \mu}=
{\kappa}^{-1}\, k^{\bar \mu}$. If we rewrite Eq.~(\ref{eq:Kint}) in terms
of the rescaled coordinates, we obtain
\begin{eqnarray}
\int_{\hat {V_0}}^{\infty}\, T_{\hat \mu \hat \nu} \, k^{\hat \mu}\,k^{\hat
\nu}
\, d{\hat V} \, & = & \, - { 1 \over {768\,\pi\,{\kappa}^2\,M^2\,\hat {V_0} } }
                                          \nonumber \\
& = & \, - { 1 \over {48 \pi \, \hat {V_0} } } \,.     \label{eq:Kinthat}
\end{eqnarray}

Compare this result with our result for null rays outside the horizon, i.e.,
Eq.~(\ref{eq:anec/nint}), in which ${\delta \lambda}$ is interpreted
as the characteristic affine parameter
distance along the null geodesic over which the negative energy
density falls off. Let us now show that a similar interpretation holds
{\it on} the horizon. The integrand of Eq.~(\ref{eq:Kinthat}) is
equal to $- 1/(48\,\pi\,{\hat V}^2)$, so
\begin{eqnarray}
\int_{\hat {V_0}}^{\infty}\, T_{\hat \mu \hat \nu} \, k^{\hat \mu}\,k^{\hat
\nu}
\, d{\hat V} \,& = & \, -
\frac{1}{48\pi} \int_{\hat {V_0}}^{\infty}\, \Biggl( {1 \over { {\hat V}^2} }
\Biggr)\, d{\hat V}                                        \nonumber \\
& = & \, - { 1 \over {48 \pi \, \hat {V_0} } } \,.     \label{eq:Kinthat/2}
\end{eqnarray}
Note that $\hat {V_0}$ can be interpreted as the characteristic affine
parameter distance over which the integrand falls off, i.e., as  $\hat V$
increases from $\hat V = \hat {V_0}$ to $\hat V = 2 \hat {V_0}$,
$T_{\hat \mu \hat \nu} \, k^{\hat \mu}\,k^{\hat \nu}$ falls off to $1/4$ of
its initial value. Thus the interpretation of the QI for null geodesics
on the horizon is the same as that for Eq.~(\ref{eq:anec/nint}).

\section{4D Evaporating Black Holes}
\label{sec:4DBH}
      In this section, we examine AWEC and ANEC in four-dimensional
evaporating black hole spacetimes. In these spacetimes, the components
of the renormalized expectation value of the stress-energy tensor are
only known numerically. Accordingly, we make use of the numerical data
of Elster \cite{ELSTER} for the conformally-coupled massless scalar field
and that of Jensen, McLaughlin, and Ottewill \cite{JMO}
for the electromagnetic field, evaluated in the Unruh vacuum state.
We examine the cases of outgoing radial
timelike observers, outgoing radial null geodesics, and orbiting null
geodesics. Only geodesics outside the horizon are considered, as no
numerical data are available for $r \leq 2M$.
We find that although AWEC and ANEC are violated in all three
cases, there appear to be QI-type bounds in each case which constrain the
degree of the violations, as in two-dimensional spacetime.

 These results
might at first sight seem to be in disagreement with recent results of
Visser \cite{VBOOK}. He found that for the renormalized vacuum expectation
value of the stress-energy tensor of a conformally-coupled massless scalar
field in Schwarzschild spacetime, ANEC is satisfied for all null geodesics
that reach spatial infinity, and for orbiting geodesics at $r=3M$. However,
his results are applicable
for the Hartle-Hawking vacuum state, which represents a black hole in
thermal equilibrium with a surrounding external thermal bath of radiation.
It is therefore not too surprising to find that ANEC holds in that case.
By contrast, our analysis is performed in the Unruh vacuum state, which
corresponds to the situation of a black hole evaporating into empty space.

\subsection{Outgoing Radial Timelike Observers}

The four-velocity of an outgoing radial timelike observer in 4D
Schwarzschild spacetime is given by
\begin{equation}
u^\mu = \Bigl( {k \over C},\, \sqrt{k^2 -C},\,0,\,0 \Bigr)\,, \label{eq:umu/4D}
\end{equation}
where again $C=(1-2M/r)$. Let us adopt the notation of Elster \cite{ELSTER}
so that: $\mu (r)=-{T^t}_t\,,\quad p_r (r)= {T^{r}}_r\,,\quad
L=-4\pi\,r^2\,C\,T_{t r}$,
where the components of the stress-tensor here represent renormalized
expectation values in the Unruh vacuum state. Therefore, we may write
\begin{equation}
T_{\mu \nu} u^{\mu} u^{\nu} = { {k^2} \over C } \, (\mu + p_r)\,- p_r \,
- { {kL} \over {2 \pi\, r^2\, C^2} } \, \sqrt{k^2\,-C}\,.   \label{Tuu/4D}
\end{equation}
If a QI-type bound of the kind found in 2D, i.e., Eq.~(\ref{eq:awec/qi}),
also holds in 4D, then we might expect it \cite{COMFORM} to have the form
\begin{equation}
\int_{0}^{\infty}\, T_{\mu\nu} {u^\mu} {u^\nu} d\tau \gprox
-{1 \over {M^2 \, \delta \tau}}\,,                   \label{eq:awec/qi-4D}
\end{equation}
where again $\delta \tau$ is the characteristic proper time which the
observer spends in the negative energy. (Note that since the
fields we will be considering are massless, the mass $M$ of the black
hole is only natural length scale which appears in the problem.) We want to
numerically compare the left and right-hand-sides of
Eq.~(\ref{eq:awec/qi-4D}), in order to see if the inequality is satisfied.

   The integrand in Eq.~(\ref{eq:awec/qi-4D})
drops off rapidly with increasing $r$ near the horizon (unlike in the
2D case), hence the dominant
contribution to the AWEC integral comes from the initial part of the
trajectory. We will therefore approximate the AWEC integral as
\begin{equation}
\int_{0}^{\infty}\, T_{\mu\nu} {u^\mu} {u^\nu} d\tau
 \approx \,  T (r)|_{r_0} \,\delta \tau  \,,  \label{eq:4DAWEC/prox}
\end{equation}
where $T (r) \equiv  T_{\mu\nu} {u^\mu} {u^\nu}$ is understood to be
evaluated at the starting point $r_0$ (minimum $r$) of the trajectory. Since
the magnitude of $T (r)$ rapidly decreases near the horizon, our estimate of
using $T (r_0)$ gives a result which is somewhat more negative
than the true result. Therefore, correction of this estimate only goes
in the direction of strengthening the inequality.

Our results for the electromagnetic field are displayed in Table 1.
The values of $\mu$ and $p_r$ for the electromagnetic field are
from the numerical data given in Ref. \cite{JMO}, and the value of
$L= 3.4 \times 10^{-4}/{M^2}$ used in this case is the photon luminosity given
by Page \cite{PAGE}. In the fourth column of Table 1, $\delta r$ is chosen
such that the magnitude of $T(r)$ decreases by at least a factor of two
from $r_0$ to $r_0 + \delta r$. The characteristic proper time, $\delta \tau$,
is computed from
\begin{equation}
\delta \tau = \int_{r_0}^{r_0 + \delta r} \frac{dr}{\sqrt{k^2 -C}} \, .
\end{equation}
The next to last column in the table is our
approximation
of the AWEC integral given in Eq.~(\ref{eq:4DAWEC/prox}), while the last
column represents the right-hand-side of our QI given by
Eq.~(\ref{eq:awec/qi-4D}). By examining the last two columns of Table 1,
we see that the QI, Eq.~(\ref{eq:awec/qi-4D}), is easily satisfied.
A similar analysis for the conformally-coupled massless scalar field
shows that the QI is easily satisfied in that case as well.
Numerical values of $\mu\,, p_r$,
and $L= 7.44 \times 10^{-5}/{M^2}$ for the scalar field were obtained
from the numerical data given in Ref. \cite{ELSTER}.
\begin{table}
\begin{tabular}{|r|r|r|r|r|r|r|r|r|} \hline
$r_0$ & $\mu(r_0) M^4$ & $p_r(r_0)M^4$ & $\delta r/M$ & $k$ &
$\delta\, \tau/M$ &
$T\,\,\delta\, \tau$ & $-(M^2 \, \delta\,\tau)^{-1}$ \\ \hline
2.1 & $-2.7\times 10^{-3}$ & $1.9\times 10^{-3}$ & 0.5
& 0.5 & $1.77$ & $-1.1\times 10^{-2}$ & -0.56 \\
& & & & 1.0 & $0.54$ & $-1.0\times 10^{-2}$ & -1.85 \\
& & & & 5.0 & 0.10 & $-4.4\times 10^{-2}$ & -10 \\
& & & & 10.0 & 0.05 & $-8.5\times 10^{-2}$ &  -20 \\
2.2 & $-1.8\times 10^{-3}$ & $1.5\times 10^{-3}$ & 1.0
& 1 & $1.16$ & $-5.7\times 10^{-3}$ & -0.86 \\
& & & & 5 & 0.20 & $-1.7\times 10^{-2}$ & -5 \\
& & & & 10 & 0.10 & $-3.5\times 10^{-2}$ & -10 \\
2.5 & $-1.1\times 10^{-3}$ & $6.8\times 10^{-4}$ & 1.0
& 1 & $1.22$ & $-3.3\times 10^{-3}$ & -0.82 \\
& & & & 5 & 0.20 & $-1.0\times 10^{-2}$ & -5 \\
& & & & 10 & 0.10 & $-2.1\times 10^{-2}$ & -10 \\
3 & $-5.4\times 10^{-4}$ & $3.2\times 10^{-4}$ & 2
& 1 & $2.82$ & $-2.8\times 10^{-3}$ & -0.45 \\
& & & & 5 & 0.40 & $-6.8\times 10^{-3}$ & -2.5 \\
& & & & 10 & 0.20 & $-1.3\times 10^{-2}$ & -5 \\ \hline
\end{tabular}
   \caption{Outgoing timelike geodesics moving through the quantized
electromagnetic field.}
\end{table}

\subsection{Radial Null Rays}
    The components of the tangent vector to an affinely parameterized
outgoing radial null geodesic (outside the horizon) are given by
\begin{equation}
K^{\mu} = E \, ( C^{-1},\, 1\,,0\,,0) \,.       \label{eq:snullv/4D}
\end{equation}
Thus we have
\begin{equation}
T_{\mu \nu}  K^{\mu}  K^{\nu}  = { {E^2} \over {2\pi\,{(r-2M)}^2 } }\,
\Biggl \{ 2\pi r\, (r-2M)\, [\mu(r) + {p_r}(r)] \,-L \Biggr \} \,.
                                                   \label{eq:TKK/4D}
\end{equation}
The 4D analog of the QI for null geodesics in 2D given by
Eq.~(\ref{eq:anec/nint}) is
\begin{equation}
\int_{0}^{\infty}\, T_{\mu\nu} {K^\mu} {K^\nu} d\lambda \gprox
-{1 \over {M^2 \, \delta \lambda}}\,,                   \label{eq:4Danec/qi}
\end{equation}
where ${\delta \lambda}$ is again interpreted as the characteristic affine
parameter distance along the null geodesic over which the negative energy
density falls off. If we define
\begin{equation}
T \equiv { 1 \over {E^2} } \, (T_{\mu\nu} {K^\mu} {K^\nu}) \,,
\label{eq:T-def}
\end{equation}
and use the fact that $dr/{d \lambda} = E$ and $\delta \lambda = {\delta r}/E$,
then it is easily seen that the QI, Eq.~(\ref{eq:4Danec/qi}) is equivalent to
\begin{equation}
\int_{r_0}^{\infty} \, T \, dr \gprox - { 1 \over {M^2 \, \delta r} } \,,
                                                \label{eq:4Danec-r}
\end{equation}
where $r_0 > 2M$ is the initial starting value of $r$ for the outgoing null
geodesic and $\delta r$ is the characteristic interval in $r$ over which
the negative energy falls off. We numerically evaluate each side of
Eq.~(\ref{eq:4Danec-r}) to see if the inequality is satisfied.
Following a similar procedure to that used for outgoing radial timelike
observers, we approximate the ANEC integral by
\begin{equation}
\int_{r_0}^{\infty} \, T \, dr \approx T(r_0) \, \delta r \,.
                                                \label{eq:4Danecprox}
\end{equation}
As discussed in the last section, this approximation gives a result which
is somewhat more negative than the true result. Therefore, if the QI holds
for our approximation, then the inequality in the actual case is even
stronger.

Our results for the scalar field are given in Table 2,
and those for the electromagnetic field appear in Table 3, where
as before the numerical data of Refs. \cite{ELSTER} and \cite{JMO},
were used for the scalar and electromagnetic fields, respectively.
The next to last column in each table is our approximation
of the ANEC integral given in Eq.~(\ref{eq:4Danecprox}), while the last
column represents the right-hand-side of our QI given by
Eq.~(\ref{eq:4Danec-r}). By examining the last two columns of each table,
we see that the QI, Eq.~(\ref{eq:4Danec-r}), is easily satisfied and
hence so is Eq.~(\ref{eq:4Danec/qi}). Note that the inequalities in 4D
seem to be satisfied by a wider margin than the corresponding QI's in 2D.
This is due to the fact that the numerical factors which appear in the
components of the renormalized expectation values of the stress-tensor
are characteristically smaller in 4D than in 2D spacetime, at least for radial
null geodesics in the scalar field case.
\begin{table}
\begin{tabular}{|r|r|r|r|r|r|} \hline
$r_0/M$ & $\mu(r_0)M^4$ & $p_r(r_0)M^4$
& $\delta r/M$ & $T\, \delta r\, M^3$ & $- M/\delta r$ \\ \hline
2.5 & $-4\times 10^{-6}$ & 0 & 0.5 & $-3.4\times 10^{-5}$ & -2 \\
4 & $-4\times 10^{-7}$ & $-3\times 10^{-7}$ & 3 & $-4.8\times 10^{-6}$ & -0.3
\\
5 & $-1.5\times 10^{-6}$ & $-3\times 10^{-7}$ & 5 & $-7.8\times 10^{-6}$
&-0.2\\
   \hline
\end{tabular}
  \caption{Data is given for outgoing null geodesics, where $\mu$ and $p_r$ are
the energy density and pressure for a massless scalar field.}
\end{table}

\begin{table}
\begin{tabular}{|r|r|r|r|r|r|} \hline
$r_0/M$ & $\mu(r_0) M^4$ &
$p_r(r_0) M^4$
& $\delta r/M$ & $T\, \delta r\, M^3$ & $- M/\delta r$ \\ \hline
2.1 & $-2.7\times 10^{-3}$ & $1.9\times 10^{-3}$ & 0.5 & -0.01 & -2 \\
2.2 & $-1.8\times 10^{-3}$ & $1.5\times 10^{-3}$ & 1 & -0.004 & -1 \\
2.5 & $-1.1\times 10^{-3}$ & $6.8\times 10^{-4}$ & 1 & -0.002 & -1 \\
3 &$-5.4\times 10^{-4}$ & $3.2\times 10^{-4}$ &
                               2 & $-1.4\times 10^{-3}$& -0.5 \\ \hline
\end{tabular}
  \caption{Data is given for outgoing null geodesics, where $\mu$ and $p_r$ are
the energy density and pressure for the electromagnetic field.}
\end{table}

   In our calculations, we have ignored the backreaction of the Hawking
radiation on the background spacetime. It might be better to repeat
our calculations on an evolving background, such as a Vaidya spacetime.
However, we can estimate the effects of backreaction on an outgoing
null ray just outside the horizon in the following way. Our approximation
should hold as long as the time of escape of a null ray from near the horizon
is small compared to the evaporation time of the black hole.
Here these times are measured by an observer at infinity.
For such a null ray we have
\begin{equation}
\int_{0}^{t_{esc}} \, dt\,=\, \int_{2M+\epsilon}^{xM} \,
{(1-2M/r)}^{-1} \,dr \,.
\end{equation}
Solving for $t_{esc}$ yields
\begin{equation}
t_{esc} = xM \,+ \, 2M\,{\rm ln} \Biggl({x \over 2} -1 \Biggr)\, -2M \,
- \,\epsilon \, +\, 2M\, {\rm ln} \Biggl({2M \over \epsilon }\Biggr) \,,
                                                        \label{eq:tesc}
\end{equation}
where $t_{esc}$ is the time for the ray to go from $r=2M+\epsilon$ to
$r=xM$. For $x \gg 2$ and finite, in the limit of small $\epsilon$:
\begin{equation}
t_{esc} \rightarrow 2M\, {\rm ln} \Biggl({2M \over \epsilon }\Biggr) \,.
                                                    \label{eq:lim/tesc}
\end{equation}
By contrast, the evaporation time of the hole is
$t_{evap} =  A M^3$, where $A \approx 10^4/n$, and $n$ is the number of species
of particles, each of which is assumed to contribute approximately
$10^{-4}\,M^{-2}$
to the black hole's luminosity. Our approximation should break down
only when $\epsilon \lprox 2M\, {\rm exp}[-A M^2/2]$.

\subsection{Orbiting Null Rays}
    We now examine orbiting null geodesics in Schwarzschild spacetime.
The line element is
\begin{equation}
ds^2 = -C dt^2 + C^{-1} dr^2 + r^2({d\theta}^2 + {\rm sin}^2{\theta}{d\phi}^2)
\,,                                         \label{eq:4DMetric}
\end{equation}
where again $C=(1-2M/r)$.
For an orbiting null ray in Schwarzschild with $\theta=\pi/2=const$,
\begin{equation}
{ {dt} \over {d\lambda}}= \, {r \over {\sqrt C}} \,
                          { {d\phi} \over {d\lambda} }\,.  \label{eq:dt/dlamb}
\end{equation}
Here we choose ${ {d\phi} / {d\lambda} } > 0$, for our future-directed
ray. From the geodesic equations, we have
\begin{eqnarray}
{\dot t} & \equiv & { {dt} \over {d\lambda} } ={ E \over C} \,,
\\                                              \label{eq:geod/t}
{\dot \phi} & \equiv & { {d\phi} \over {d\lambda} }  = {L \over {r^2}} \,,
                                                   \label{eq:geod/phi}
\end{eqnarray}
where $E$ and $L$ are constants. It follows that
\begin{equation}
{L \over E} = { r \, {C^{-1/2}} }\,.              \label{eq:L/E def}
\end{equation}
At $r=3M$, we have
\begin{equation}
{L \over E} \biggl|_{r=3M} = 3{\sqrt 3}\, M  \,.
                                                     \label{eq:bo/def}
\end{equation}
{}From the numerical values in Figs. (1) and (3) of Ref. \cite{JMO}
for the quantized electromagnetic field, and Eq.~(\ref{eq:bo/def}),
we find that
\begin{eqnarray}
T_{\mu \nu} K^{\mu} K^{\nu} \left|_{r=3M} \right.
& = & 3\, E^2\, \Bigl[ -T^{t}_{t}\, +
\, T^{\phi}_{\phi} \Bigr] \left|_{r=3M} \right.    \nonumber \\
& \approx &  -3.1 \times 10^{-4} \,{E^2 \over M^4} \,, \label{eq:T3M}
\end{eqnarray}
where $K^{\mu}=dx^{\mu}/{d\lambda}$ is the tangent to the null geodesic.
For one orbit, $\Delta \phi= 2\pi = (L/9M^2)\, \Delta \lambda$, so
\begin{equation}
{E \over M}= \, { {2\sqrt{3} \pi} \over {\Delta \lambda} }\,.
\end{equation}
We may therefore rewrite Eq.~(\ref{eq:T3M}) as
\begin{equation}
T_{\mu \nu} K^{\mu} K^{\nu}|_{r=3M}
\approx \,  -3.7 \times 10^{-2} \,{ 1 \over {M^2 \, {(\Delta \lambda)}^2}} \,.
\end{equation}
Integration of $T_{\mu \nu} K^{\mu} K^{\nu}$ over one orbit gives
\begin{equation}
\int_{0}^{\Delta \lambda}
\,{T_{\mu\nu} {K^\mu} {K^\nu} d\lambda}
 \approx   -3.7 \times 10^{-2} \,{ 1 \over {M^2 \, {(\Delta \lambda)}}}\, .
                                              \label{eq:ANEC-1norbit}
\end{equation}
Thus we see that for orbiting null geodesics
\begin{equation}
\int_{0}^{\Delta \lambda}
\,{T_{\mu\nu} {K^\mu} {K^\nu} d\lambda} \gprox
\, -{ 1 \over {M^2 \, {(\Delta \lambda)}}}\, .
\label{eq:ANECINQ-1norbit}
\end{equation}

The expressions given by Eqs.~(\ref{eq:ANEC-1norbit}) and
{}~(\ref{eq:ANECINQ-1norbit}) are of the same form as those for radial null
geodesics discussed earlier. The difference here is that we integrate only
over one orbit, i.e., we do not consider multiple traversals through the
same negative energy region. The inclusion of the contributions of an
arbitrary number of orbits would violate the inequality given in Eq.
{}~(\ref{eq:ANECINQ-1norbit}). (Since the $r=3M$ orbit
is unstable, realistically we would expect the null ray to at best only make
a few orbits.) However, it is of interest to note that if we
consider the null geodesics which generate the boundary of the future of
an event $P$ with $r=3M$, then the orbiting null geodesics leave this
boundary after half an orbit. This is due to the fact that orbiting null
geodesics originating at $P$, which initially set off in opposite spatial
directions, encounter a conjugate point (i.e., the geodesics cross) on the
far side of the black hole after half an orbit \cite{BORDE-PRIVATE},
\cite{PENROSE-TDTR}.
At this point, the geodesics leave the boundary of the future of $P$ (i.e.,
an event $Q$ which lies beyond the conjugate point along either null geodesic
can be connected to $P$ by a timelike curve), and therefore are not achronal.

This observation may be relevant
in the context of proving singularity theorems in the presence of negative
energy. For example, in the proof of a singularity theorem such as the Penrose
theorem, one uses either the local or the averaged null energy condition to
insure the focusing of the null geodesics which generate the boundary of the
future of a closed trapped surface. Geodesics of the orbiting variety
considered here would not relevant, at least in the spherically symmetric case,
since they leave the boundary of the future of a point in any case,
irrespective of energy condition-focusing arguments. We also note in passing
that, at least in this simple case, while
these null geodesics are in the boundary they obey the inequality, Eq.
{}~(\ref{eq:ANECINQ-1norbit}).

Let us compare the previous results with the following situation. Consider
a 4D Minkowski spacetime with one spatial dimension compactified,
and a quantized, minimally-coupled, massless scalar field in the Casimir vacuum
state on this background.  Let $K^{\mu}= E(1,1,0,0)$ be
the tangent to a null geodesic in the $x$-direction. The vacuum
expectation value of the stress-tensor in this state is
\begin{equation}
T_{\mu\nu} =\, { {\pi}^2 \over {45\, L^4} }\, \times
{\rm diag} \, \{-1,-3,1,1\} \,,                         \label{eq:T/Cas}
\end{equation}
where we have taken $x$ as the compactified dimension with circumference $L$.
It follows that
\begin{equation}
T_{\mu \nu} K^{\mu} K^{\nu}=\, -{ {4{\pi}^2 E^2} \over {45\, L^4} }\,.
                                                          \label{eq:tmunu}
\end{equation}
The affine parameter lapse for one orbit is $\lambda_p = L/E$.
If we integrate over one orbit, we find that
\begin{eqnarray}
\int_{0}^{\lambda_p}
\,{T_{\mu\nu} {K^\mu} {K^\nu} d\lambda} & = &
\Bigl( T_{\mu\nu} {K^\mu} {K^\nu} \Bigr)\, (\lambda_p)  \nonumber \\
& = & -{ {4{\pi}^2} \over
{45\, L^2 \, \lambda_p} } \, .
                                              \label{eq:AWEC-1orbit/4DCas}
\end{eqnarray}
We see that this result has the same form as Eq.~(\ref{eq:ANEC-1norbit})
for the black hole case. Both of these expressions are
invariant under rescaling of the affine parameter (as is
Eq.~(\ref{eq:4Danec/qi}) for radial null geodesics). Here, as in the black hole
case, we have ignored the possibility of multiple orbits, which would violate
the inequality. However, it is possible that in the 4D compactified case, as
was found previously in 2D \cite{FR95}, an ANEC-type bound might hold on the
{\it difference} between the expectation values of
$T_{\mu\nu} {K^\mu} {K^\nu}$
in an arbitrary quantum state and in the Casimir vacuum state.

\section{Conclusions}
\label{sec:CONC}
  In this paper, we have examined AWEC and ANEC in both two and four-
dimensional evaporating Schwarzschild black hole spacetimes, for
quantized massless scalar and electromagnetic fields. Our 2D results are the
following: It was
discovered that AWEC is satisfied for outgoing radial timelike geodesics
which start just outside the horizon and reach infinity, for the Unruh vacuum
state. This is due to the positive contribution to the AWEC integral of
the Hawking radiation at large distances, which outweighs the
contribution of the negative energy near the horizon.  By contrast, AWEC is
violated for such geodesics in the Boulware vacuum state. Similarly, ANEC
is violated for half-complete outgoing radial null geodesics, which start just
outside the horizon, in both the Unruh and Boulware vacuum states. This is
also true on the horizon in the Unruh vacuum state. However, ANEC is satisfied
along ingoing null geodesics and along outgoing null geodesics inside the
horizon in the Unruh vacuum state, due to a large positive contribution to the
integral near the singularity.

In 4D, we examined AWEC and ANEC in the Unruh vacuum state, using previously
calculated numerical values of the stress-tensor components. There it was
found that AWEC is violated for outgoing timelike geodesics which start
just outside the horizon and reach infinity. (Note that in 4D, the Hawking
radiation decays like $1/r^2$, so the large positive contribution
to the AWEC integral found in
2D does not arise here.) Similarly, we found that ANEC is violated for
analogous outgoing null geodesics. This condition is violated
for orbiting null geodesics as well.

However, in all
cases where the conditions are violated, there appear to be QI-type
bounds which limit the degree of the
violations. Let $\delta \tau$ be the
characteristic proper time for timelike geodesics,
and $\delta \lambda$ be the affine parameter
length for null geodesics, over which the negative energy density
changes significantly. In 2D the QI bounds have the form that the integrated
negative energy density along the timelike or null geodesic (i.e., the
AWEC or ANEC integral) is greater than or equal to minus the
inverse of $\delta \tau$ or $\delta \lambda$, respectively. In 4D, the
corresponding inequalities have additional factors of $M^{-2}$ on the
right-hand-sides.

Can the semi-classical effects of negative energy, in processes such as
black hole evaporation, invalidate the singularity theorems before
quantum gravity effects become significant? Although we do not have a
definite proof in the general case, our results suggest that the answer
may be no. It remains possible that even though ANEC (or even a weaker
energy condition \cite{Y95}) might fail in some regions of a given spacetime,
it may hold in enough other regions for the conclusions of the singularity
theorems to be maintained. Our (admittedly) 2D results for ANEC would seem
to support this view.
In Ref. \cite{TR-88}, it was shown that if ANEC is satisfied along the
null geodesics which generate the boundary of the future
of a trapped surface, then Penrose's singularity theorem will still hold.
There the local null energy condition was replaced by ANEC along half-complete
null geodesics.
However, two points should be emphasized about that result. First, it is
not necessary to assume that the ANEC integral is non-negative to prove the
result \cite{TR-88,Y95}. In fact, from Lemma 3 of Ref. \cite{Galloway} or
Eq. (5) of Ref. \cite{TR-88} and the Einstein equations, it is sufficient to
have
\begin{equation}
\int_{0}^{\infty} \,T_{\mu \nu} K^{\mu} K^{\nu} \, d\lambda
\geq \, { {\theta_0} \over 2} \,,               \label{eq:W-ANEC}
\end{equation}
where $\theta_0<0$ is the initial expansion of the null geodesic congruence
at the trapped surface. The ANEC integral may be negative as long as it is
not more negative than $\theta_0/2$. Physically this implies that a
singularity will still form provided that the defocusing effects
due to the presence of any negative energy are more than
compensated for by the initial convergence of the null geodesics produced
by the trapped surface.

Second, the existence of only {\it one} such trapped surface whose orthogonal
null geodesics obey either ANEC or the weaker inequality,
Eq.~(\ref{eq:W-ANEC}), is required to prove the occurrence of a singularity.
There may exist other trapped surfaces in the spacetime whose orthogonal null
geodesics do not obey either of the inequalities. However, as long as there
exists {\it at least one} trapped surface with the desired properties,
a singularity is inevitable.

Consider an evaporating black hole spacetime, with backreaction included.
The Penrose diagram for the standard scenario is shown in Fig. (3).
Let us assume that this standard picture is correct. The line $H^{+}$ is the
event horizon, the dashed curve $A$ is the apparent horizon, and the
lines $g_1$, $g_2$, and $g_3$ are outgoing null geodesics
orthogonal to the trapped surfaces labelled $1$, $2$, and $3$, respectively.
The outgoing null rays from $1$ do not focus but instead reach future
null infinity, so one expects that {\it any} averaged energy
condition which would
insure focusing, such as ANEC or Eq.~(\ref{eq:W-ANEC}), would fail along them.
The closest analogs to these rays in our analysis are the outgoing null rays
just outside the horizon shown in Fig. (1). It was found that in both
two and four-dimensional Schwarzschild spacetime, ANEC was violated along
these rays, although in each case a QI limits the degree of the violation.
One would expect that if an analogous QI
holds in more general cases, such as Fig. (3), it will not be strong
enough to guarantee focusing of null rays in the region between $H^{+}$ and $A$
in Fig. (3). We could not examine the weaker inequality Eq.~(\ref{eq:W-ANEC}),
since in a static evaporating black hole background there are no trapped
surfaces outside the horizon. For the outgoing null rays from $2$ in Fig. (3)
(a trapped surface inside the collapsing matter), one would suspect that
ANEC probably holds. In this case, the focusing effects of the matter of
the star most likely overwhelm any defocusing by negative energy.
For the outgoing null rays from $3$ (a trapped surface
``behind'' the event horizon, in the vacuum outside the collapsing matter),
one might expect either ANEC or a similar but  weaker inequality such as
Eq.~(\ref{eq:W-ANEC}) to hold, since the
null rays eventually do run into a singularity. This is assuming, of course,
that the ``standard'' scenario depicted in Fig. (3) is in fact the correct
one - a view which could well be wrong. In our 2D analysis, we found that ANEC
was satisfied along these rays, and also along ingoing null rays.

   We showed that our approximation of neglecting backreaction should fail
only when the time of escape of a null ray from near the horizon
is comparable to the evaporation time of the black hole. This
implied that such a ray would have to start out extremely close to the
horizon. Nevertheless, it would be useful to attempt a similar analysis
to ours for evaporating black hole spacetimes with backreaction taken
into account. The situation is complicated by our ignorance of the
detailed form of the stress-tensor components in such cases.

Recent work of Kuo and Ford \cite{KF,Kuo} has shown that negative energy
densities in flat spacetime are subject to large fluctuations. In these
circumstances, one does not expect the semi-classical Einstein equations
to be a good approximation. If the same
is true of negative energy densities in curved spacetime, and we see no reason
to believe otherwise, then the question arises as to whether such
fluctuations could cause a gross failure of the semi-classical approximation
at scales well above the Planck scale. Large stress-tensor fluctuations would
presumably produce large fluctuations in spacetime geometry. Therefore,
it is very important to determine whether such fluctuations might render
suspect the semi-classical picture of the Hawking process, or the
conclusions of the singularity theorems in the presence of negative energy.
On the other hand it could turn out, for example in the Hawking process,
that if the timescale of the stress-tensor fluctuations is very short
compared to other relevant time scales, such as the evaporation time
of the hole, then the standard analysis may still be valid. In the proofs
of the singularity theorems, one assumes a fixed classical spacetime
background which determines the causal structure. If the background geometry
fluctuates, it is not clear that the usual global
techniques employed in classical general relativity will still be useful
\cite{BORDE-PRIVATE2}. These questions are currently under active
investigation.

\vskip 0.2 in
\centerline{\bf Acknowledgements}
The authors would like to thank Arvind Borde, Eanna Flanagan, Matt Visser,
Bob Wald, and Ulvi Yurtsever for helpful discussions. TAR would like to thank
the members of the Tufts Institute of Cosmology for their kind hospitality
and encouragement while this work was being done.
This research was supported in part by NSF Grant No. PHY-9208805 (Tufts), by
the U.S. Department of Energy (D.O.E.) under cooperative agreement
\# DF-FC02-94ER40818 (MIT), and by a CCSU/AAUP Faculty Research Grant.

\newpage
\section*{Figure Captions}
\begin{itemize}

\item{[1]} Radial null geodesics in the spacetime of a black hole formed by
gravitational collapse: an ingoing geodesic (a), an outgoing geodesic inside
the horizon (b), and an outgoing geodesic outside of the horizon (c).

\item{[2]} The spacetime of a black hole formed by gravitational collapse.
The null ray which lies in the future horizon enters the collapsing body
(shaded region), and then re-emerges into the vacuum spacetime at $V_{min}$.

\item{[3]} The spacetime of an evaporating black hole, including the effects of
backreaction.

\end{itemize}


\begin{thebibliography}{--}

\bibitem{HE}  S.W. Hawking and G.F.R. Ellis, {\it The Large Scale
              Structure of Spacetime} (Cambridge University
              Press, London, 1973), pp. 88-96.

\bibitem{EGJ} H. Epstein, V. Glaser, and A. Jaffe, Nuovo Cim. {\bf
              36}, 1016 (1965).


\bibitem{C} H.B.G. Casimir, Proc. Kon. Ned. Akad. Wet.  {\bf B51},
793 (1948).

\bibitem{T} F.J. Tipler, Phys. Rev. D {\bf 17}, 2521 (1978).

\bibitem{C-E} C. Chicone and P. Ehrlich, Manuscr. Math. {\bf 31}, 297 (1980).

\bibitem{Galloway} G.J. Galloway, Manuscripta Math. {\bf 35}, 209 (1981).

\bibitem{B87} A. Borde, Class. Quantum Grav. {\bf 4}, 343 (1987).

\bibitem{TR-86} T.A. Roman, Phys. Rev. D {\bf 33}, 3526 (1986).

\bibitem{TR-88} T.A. Roman, Phys. Rev. D {\bf 37}, 546 (1988).

\bibitem{FSW} J. Friedman, K. Schleich, and D. Witt, Phys. Rev. Lett.
               {\bf 71}, 1486 (1993).


\bibitem{CONCOM}  Note that in a general spacetime, unlike the relation
between the local null and weak energy conditions, ANEC does not simply
follow by continuity from AWEC. An example of this is illustrated in
Sec.~(\ref{sec:2DEVAP}) for the Unruh vacuum state.


\bibitem{K} G. Klinkhammer, Phys. Rev. D {\bf 43}, 2542 (1991).

\bibitem{Folacci} A. Folacci, Phys. Rev. D {\bf 46}, 2726 (1992).
\bibitem{WY} R. Wald and U. Yurtsever, Phys. Rev D {\bf 44}, 403
(1991).

\bibitem{VISSER} M. Visser, Phys. Lett. {\bf B349}, 443 (1995).

\bibitem{Y94} U. Yurtsever, Phys. Rev. D {\bf 51}, 5797 (1995).

\bibitem{Y95} U. Yurtsever, Phys. Rev. D {\bf 52}, R564 (1995).

\bibitem{MT} M. Morris and K. Thorne, Am. J. Phys. {\bf 56},
               395 (1988).

\bibitem{MTY} M. Morris, K. Thorne, and U. Yurtsever, Phys. Rev.
               Lett. {\bf 61}, 1446 (1988).

\bibitem{FR95} L. Ford and T. Roman, Phys. Rev. D, {\bf 51}, 4277 (1995).


\bibitem{F78} L.H. Ford, Proc. Roy. Soc. Lond. A {\bf 364}, 227
               (1978).

\bibitem{F91} L.H. Ford, Phys. Rev. D {\bf43}, 3972 (1991).

\bibitem{CSAMPLE} This particular choice was initially made in Ref. \cite{F91}
purely for mathematical
convenience. Presumably, one could prove analogous inequalities with other
choices of sampling functions.

\bibitem{FR90} L.H. Ford and T.A. Roman, Phys. Rev. D {\bf 41}, 3662
               (1990).

\bibitem{FR92} L.H. Ford and T.A. Roman, Phys. Rev. D {\bf 46}, 1328
               (1992).

\bibitem{FR-Worm} L.H. Ford and T.A. Roman, {\it Quantum Field Theory
Constrains Traversable Wormhole Geometries}, manuscript in preparation.

\bibitem{Y} U. Yurtsever, Class. Quantum Grav. {\bf 7}, L251 (1990).


\bibitem{P65}   R. Penrose, Phys. Rev. Lett. {\bf 14}, 57 (1965).


\bibitem{H75} S. W. Hawking, Commun. Math. Phys. {\bf 43}, 199 (1975).

\bibitem{B-TC}   A. Borde, ``Topology Change in Classical General Relativity",
Tufts Institute of Cosmology preprint TUTP 94-13, gr-qc 9406053.


\bibitem{FR93} L.H. Ford and T.A. Roman, Phys. Rev. D {\bf48}, 776 (1993).


\bibitem{Unruh77} W.G. Unruh, Phys. Rev. D {\bf 15}, 365 (1977).

\bibitem{2TS} Note that this expression differs somewhat in form from
Eq.~(\ref{eq:TLQI2DUNCP}). However, the latter is an inequality valid for
all quantum states in 2D Minkowski spacetime, whereas the integral in
Eq.~(\ref{eq:intT}) is evaluated for a restricted class of geodesics in a
specific quantum state in Schwarzschild spacetime. Note also that the
right-hand-side of Eq.~(\ref{eq:intT}) is more negative than the
right-hand-side of Eq.~(\ref{eq:TLQI2DUNCP}). This is presumably due to the
integration over a half-complete geodesic in Eq.~(\ref{eq:intT}); even in
Minkowski spacetime, there exist states for which Eq.~(\ref{eq:TLQI2DUNCP})
would not be true if the integration were taken only over a half-geodesic.

\bibitem{DU77} P.C.W. Davies and W.G. Unruh, Proc. Roy. Soc. Lond. A
{\bf 356}, 259 (1977).


\bibitem{WALDGR} R. M. Wald, {\it General Relativity} (Univ. of Chicago Press,
Chicago, 1984).


\bibitem{WALDQFT} R. M. Wald, {\it Quantum Field Theory in Curved Spacetime
and Black Hole Thermodynamics} (Univ. of Chicago Press, Chicago, 1994).


\bibitem{Y-PRIVATE} U. Yurtsever, private communication.


\bibitem{ELSTER} T. Elster, Phys. Lett. {\bf 94A}, 205 (1983).


\bibitem{JMO} B.P. Jensen, J.G. McLaughlin, and A.C. Ottewill, Phys. Rev.
              D {\bf 43}, 4142 (1991).

\bibitem{VBOOK} M. Visser, {\it Lorentzian Wormholes - from Einstein
to Hawking} (American Institute of Physics Press, New York, 1995),
Sec. 12.3.6, pp. 128-9.

\bibitem{COMFORM} The form of Eq.~(\ref{eq:awec/qi-4D}) is motivated by the
following considerations: If we imagine a sequence of observers shot outward
with increasing values of $k$, then $T_{\mu\nu} {u^\mu} {u^\nu} \propto k^2$
and $\delta \tau \propto k^{-1}$, so
$\int_{}^{}\, T_{\mu\nu} {u^\mu} {u^\nu} d\tau \propto (k^2)\,k^{-1} \,
\propto {(\delta \tau)}^{-1}$. The factor of $M^{-2}$ is required in order
for the dimensions to be correct. This form is also suggested by
the form of the inequality found for orbiting null geodesics,
Eq.~(\ref{eq:ANECINQ-1norbit}).

\bibitem{PAGE} D. N. Page, Phys. Rev. {\bf D13}, 198 (1976).


\bibitem{BORDE-PRIVATE} A. Borde, private communication.


\bibitem{PENROSE-TDTR} See p. 56, footnote 2 of R. Penrose,
{\it Techniques of Differential Topology in Relativity} (Society for
Industrial and Applied Mathematics, Philadelphia, 1972), for an interesting
example with timelike geodesics.

\bibitem{KF} C.-I Kuo and L.H. Ford, Phys. Rev. D {\bf 47}, 4510 (1993).

\bibitem{Kuo} C.-I Kuo, {\it Quantum Fluctuations and Semiclassical
              Gravity Theory}, Ph.D. thesis, Tufts University (1994),
              unpublished.


\bibitem{BORDE-PRIVATE2} The authors are grateful to A. Borde for this
remark.

\end{thebibliography}
\end{document}